\documentclass[aps,prl,twocolumn,showpacs]{revtex4}

\usepackage{amssymb}
\usepackage{amsmath}
\usepackage[xdvi,dvips]{graphicx}

\newcommand{\br}{{\bf r}}
\newcommand{\prt}{\partial}
\newcommand{\bu}{{\bf u}}

\begin{document}

\title{Linear ``ship waves'' generated in stationary flow of a
Bose-Einstein condensate past an obstacle}

\author{Yu.G. Gladush$^1$, G.A. El$^2$, A. Gammal$^3$, and A.M. Kamchatnov$^1$  \\
$^1$Institute of Spectroscopy, Russian Academy of Sciences,
Troitsk 142190, Moscow Region, Russia\\
$^2$ Department of
Mathematical Sciences, Loughborough University,
Loughborough LE11 3TU, UK \\
$^3$ Instituto de F\'{\i}sica, Universidade de S\~{a}o Paulo,
05315-970, C.P. 66318 S\~{a}o Paulo, Brazil
 }

\begin{abstract}
Using stationary solutions of the linearized two-dimensional
Gross-Pitaevskii equation, we describe the ``ship wave'' pattern
occurring in the supersonic flow of a Bose-Einstein condensate past
an obstacle. It is shown that these ``ship waves'' are generated
outside the Mach cone. The developed analytical theory is confirmed
by numerical simulations of the flow past body problem in the frame
of the full non-stationary Gross-Pitaevskii equation.
\end{abstract}

\pacs{03.75.Kk}

\maketitle

\section{Introduction}

Experimental creation of Bose-Einstein condensate (BEC) has led to
emergence of a new field of nonlinear wave dynamics owing to a
remarkable richness of nonlinear wave patterns supported by this
medium. First, vortices and bright and dark solitons were observed
and their dynamics was studied theoretically in framework of the
mean-field approach (see, e.g. \cite{ps} and references therein).
Then, dispersive shocks generated by a large and sharp disturbance
of BEC were found in experiment \cite{simula,hoefer} and explained
theoretically \cite{kgk04,hoefer} in framework of the Whitham theory
of modulations of nonlinear waves (see also numerical experiment in
\cite{damski}). At last, the stationary waves generated by
supersonic flow of BEC past obstacles have been recently observed
\cite{cornell}. They were studied in \cite{ek,egk1,egk2} where the
main focus was on the nonlinear component representing a train of
solitons (in the simplest case of a single soliton \cite{egk1}) or,
more precisely,  having a form of a modulated nonlinear periodic wave.
The theory developed in \cite{ek,egk1,egk2} shows that there exist
stationary spatial solutions of the Gross-Pitaevskii (GP) equation
which describe nonlinear waves supported by a supersonic BEC flow
with Mach number
\begin{equation}\label{1-1}
    M=\frac{u}{c_s}>1,
\end{equation}
where $u$ is velocity of the oncoming flow at $x\to-\infty$ and
$c_s$ is the sound speed of the long-wavelength linear waves. The
density $n$ of the condensate, as well as the components of the
velocity field, depend on the variable
\begin{equation}\label{1-2}
    w=x-ay
\end{equation}
alone, where $a$ is the slope of the phase lines with respect to the
$y$ axis and it is supposed that the velocity of the oncoming flow
is directed along the $x$ axis. Then, the Mach cone for sound waves
with infinitely large wavelength corresponds to the slope
\begin{equation}\label{2-1}
    a_M=\sqrt{M^2-1},
\end{equation}
and it was shown in \cite{egk2} that the spatial (oblique) solitons
have $a>a_M$, that is they are located inside the Mach cone. In
particular, it was shown that the shallow solitons are formed close
to the Mach cone, $a-a_M\ll a_M$, and they are asymptotically
described by the Korteweg-de Vries (KdV) equation; and deep solitons
have $a\gg1$ (i.e. they are formed at small angles with respect to
direction of the oncoming flow) and are asymptotically described by
the nonlinear Schr\"odinger (NLS) equation. On the contrary, the
linear waves are generated outside the Mach cone and they have
$a<a_M$. In fact, these linear waves had been observed in numerical
simulations some years ago \cite{WMCA} but the theory of their
generation has not been developed so far. The aim of this paper is
to develop such a theory.

\section{Linear waves generated in a BEC flow past an obstacle}

Our analysis is based on the use of the mean-filed description of
BEC dynamics in the framework of the GP equation
\begin{equation}\label{2-2}
   i\hbar\frac{\prt \psi}{\prt t}=-\frac{\hbar^2}{2m}\Delta\psi+U(\br)\psi
   +g|\psi|^2\psi,
\end{equation}
where $\psi(\br)$ is the order parameter (``condensate wave
function''), $U(\br)$ is the potential which confines atoms of a
Bose gas in a trap and/or describes interaction of the BEC with the
obstacle, and $g$ is an effective coupling constant arising due to
inter-atomic collisions with the $s$-wave scattering length $a_s$,
\begin{equation}\label{2-2}
   g=4\pi\hbar^2a_s/m,
\end{equation}
$m$ being the atomic mass. Here we consider the Bose-Einstein condensate with repulsive
interaction between particles for which $g>0$.

As suggested by the experimental set-up \cite{cornell}, we consider
a two-dimensional flow of a condensate, so that the condensate wave
function $\psi$ depends on only two spatial coordinates,
$\br=(x,y)$. To simplify the theory, we assume that the
characteristic size of the obstacle is much less than its distance
from the center of the trap, so that the oncoming flow can be
considered as uniform with constant density $n_0$ of atoms and
constant velocity ${\bf u}_0$ directed parallel to $x$ axis (see
also estimates in \cite{egk1}). It is convenient to transform
Eq.~(\ref{2-2}) to a ``hydrodynamic'' form by means of the
substitution
\begin{equation}\label{2-3}
   \psi(\br,t)=\sqrt{n(\br,t)}\exp\left(\frac{i}{\hbar}
   \int^{\br}{\bf u}(\br',t)d\br'\right),
\end{equation}
where $n(\br,t)$ is density of atoms in BEC and ${\bf u}(\br,t)$
denotes its velocity field, and to introduce dimensionless variables
\begin{equation}\label{2-4}
\begin{split}
    \tilde{x}=x/\sqrt{2}\xi,\quad \tilde{y}=y/\sqrt{2}\xi,\quad
   \tilde{t}=(c_s/2\sqrt{2}\xi)t,\\
   \tilde{n}=n/n_0,\quad \tilde{\bf u}={\bf u}/c_s,
   \end{split}
\end{equation}
where $\xi=\hbar/\sqrt{2mn_0g}$ is the BEC healing length and
numerical constants are introduced for future convenience. As a result of
this transformation we obtain the system (we omit tildes for convenience
of the notation)
\begin{equation}\label{2-5}
\begin{split}
   \tfrac12 n_t+\nabla(n\bu)=0,\\
   \tfrac12 \bu_t+(\bu\nabla)\bu+\nabla n+\nabla\left[\frac{(\nabla n)^2}{8n^2}
   -\frac{\Delta n}{4n}\right]=0,
   \end{split}
\end{equation}
where $\nabla=(\prt_x,\prt_y)$. Since we shall consider waves far
enough from the obstacle, the potential is omitted in (\ref{2-5}).

We are interested in linear waves propagating on the background flow
with $n=1$, $u=M$, $v=0$. Hence, we introduce
\begin{equation}\label{2-6}
    n=1+n_1,\quad u=M+u_1,\quad v=v_1,
\end{equation}
and linearize the system (\ref{2-5}) with respect to small deviations
$n_1,u_1,v_1$. As a result we obtain the system
\begin{equation}\label{3-1}
    \begin{split}
    \tfrac12 n_{1,t}+u_{1,x}+Mn_{1,x}+v_{1,y}=0,\\
    \tfrac12 u_{1,t}+Mu_{1,x}+n_{1,x}-\tfrac14(n_{1,xxx}+n_{1,xyy})=0,\\
    \tfrac12 v_{1,t}+Mv_{1,x}+n_{1,y}-\tfrac14(n_{1,xxy}+n_{1,yyy})=0,
    \end{split}
\end{equation}
which describes propagation of linear waves in BEC with a uniform
flow. We obtain the applicability condition of these equations, if
we notice that in the linear wave $u_1\sim Mn_1$ and the nonlinear
terms of the order of magnitude $u_1\nabla u_1\sim \nabla (Mu_1)^2$
can be neglected as long as they are much less than the linear ones
$\sim\nabla n_1$. Thus, we get the criterion
\begin{equation}\label{10a}
    n_1\ll 1/M^2.
\end{equation}
Hence, if $M$ is large enough, the linear theory is applicable to
description of waves outside the Mach cone far enough from the
obstacle, where the density amplitude $n_1$ of the wave satisfies
the condition (\ref{10a}).
For harmonic waves
$n_1,u_1,v_1\propto\exp[i(k_xx+k_yy)-i\omega t]$
the system (\ref{3-1}) yields at once the dispersion relation
\begin{equation}\label{3-2}
    \frac{\omega}2=Mk_x\pm k\sqrt{1+\frac{k^2}4},
\end{equation}
where $k=\sqrt{k_x^2+k_y^2}$.

Now we consider the stationary wave patterns far enough from the
obstacle where the condition (\ref{10a}) is supposed to be
fulfilled. In fact, this problem is analogous to the Kelvin theory
of ship waves generated by a ship moving in a deep water, but with a
different dispersion law (\ref{3-2}). Hence, we shall follow
Kelvin's method in its form presented in \cite{whitham,johnson}.

First, we notice that in a stationary wave pattern $\omega=0$ and,
hence, the components of the wavevector $\mathbf{k}=(k_x,k_y)$ are
the functions of the space coordinates $(x,y)$ connected with each
other by the relationship
\begin{equation}\label{3-3}
    G(k_x,k_y)\equiv Mk_x+k\sqrt{1+\frac{k^2}4}=0,
\end{equation}
where we have taken into account that for chosen geometry of the
BEC flow the wave must propagate upwind i.e. $k_x<0$.

Next, the ``ship wave'' pattern corresponds to a modulated
two-dimensional wave where the wavevector $\mathbf{k}$ is a gradient
of the phase \cite{whitham},
\begin{equation}\label{3-4}
    \theta=\int_0^{\br}\mathbf{k}\cdot d\br.
\end{equation}
Hence, the components $(k_x,k_y)$ satisfy the condition
\begin{equation}\label{3-5}
    \frac{\prt k_x}{\prt y}-\frac{\prt k_y}{\prt x}=0,
\end{equation}
which, with an account of (\ref{3-3}), yields the equation for $k_y$
\begin{equation}\label{3-6}
    \frac{\prt k_y}{\prt x}-f'(k_y)\frac{\prt k_y}{\prt y}=0,
\end{equation}
where $f'(k_y)$ is defined by the derivative of an implicit function
(\ref{3-3}):
\begin{equation}\label{3-7}
    f'=-\frac{\prt G/\prt k_y}{\prt G/\prt k_x}.
\end{equation}
 It follows from Eq.~(\ref{3-6}) that $k_x$ and $k_y$ are constant
along the characteristics defined as solutions of the equation
\begin{equation}\label{4-1}
    \frac{dy}{dx}=-f'(k_y).
\end{equation}

At last, since at large distances from the obstacle, the latter can be considered
as a point-like source of waves, the resulting stationary wave is centered,
that is the characteristics intersect at the point $(x,y)=(0,0)$ of the
location of the obstacle. Hence, we obtain the solution
\begin{equation}\label{4-2}
    \frac{y}x=\tan\chi=-\frac{\prt G/\prt k_y}{\prt G/\prt k_x},
\end{equation}
where $\chi$ is the angle between the $x$ axis and the radius-vector
$\br$ of the point $A$ with wavevector $(k_x,k_y)$; see Fig.~1 where
other convenient parameters are also defined, namely, the angle
$\eta$ between the wavevector $\mathbf{k}$ and a negative direction
of the $x$ axis, and the angle $\mu=\pi-\chi-\eta$ between vectors
$\mathbf{k}$ and $\br$.
\begin{figure}[bt]
\includegraphics[width=6cm,height=6cm,clip]{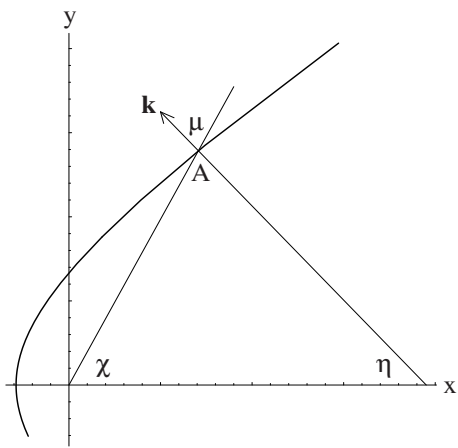}
\caption{Sketch of the wave front with wavevector $\mathbf{k}$
normal to it at the point $A$ and coordinates $\chi$ and $\eta$
defining the parameters of the wave pattern at this point.  }
\label{fig1}
\end{figure}

\begin{figure}[bt]
\includegraphics[width=8cm,height=6cm,clip]{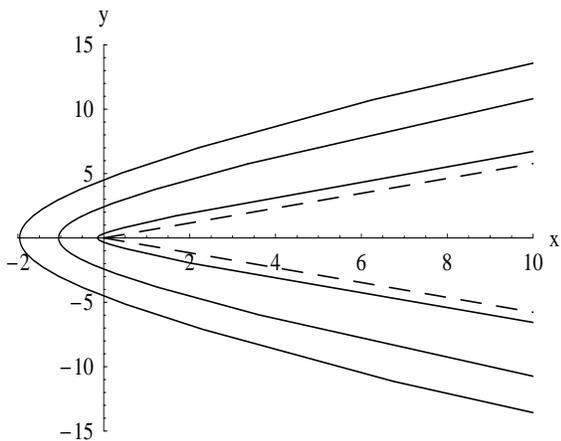}
\caption{Wave pattern of stationary linear waves generated in the
flow of BEC past a point-like obstacle. Dashed line corresponds to
the Mach cone of linear waves in the long wavelength limit. }
\label{fig2}
\end{figure}

Thus, we have
\begin{equation}\label{4-3}
    k_x=-k\cos\eta,\quad k_y=k\sin\eta,
\end{equation}
\begin{equation}\label{4-4}
    G(k,\eta)=-Mk\cos\eta+k\sqrt{1+\frac{k^2}4}=0.
\end{equation}
Then elementary calculation gives for (\ref{4-2}) the expression
\begin{equation}\label{4-5}
    \tan\chi=\frac{(2M^2-1-\tan^2\eta)\tan\eta}{(M^2+1)\tan^2\eta-(M^2-1)}
\end{equation}
and Eq.~(\ref{4-4}) yields
\begin{equation}\label{4-6}
    k=2\sqrt{M^2\cos^2\eta-1}.
\end{equation}
Thus, we have found that for a fixed value of $\eta$ the components
$(k_x,k_y)$ are constant along the line $\chi=\mathrm{const}$ with
$\chi$ defined by Eq.~(\ref{4-5}) and the length of the wavevector
given by (\ref{4-6}). Therefore the phase (\ref{3-4}) can be
conveniently calculated by integration along the line
$\chi=\mathrm{const}$ with constant vector $\mathbf{k}$, so that
\begin{equation}\label{5-1}
    \theta=(k\cos\mu)r.
\end{equation}
It means that the lines of constant phase (e.g. the wave crests)
$\theta$ are determined in parametrical form by Eq.~(\ref{4-5}) and
\begin{equation}\label{5-2}
    r=\frac{\theta}{k\cos\mu},
\end{equation}
where $k$ is given by Eq.~(\ref{4-6}) and $\mu$ can be calculated
from $\tan\mu=-\tan(\chi+\eta)$ which gives, after elementary
algebra, the expression
\begin{equation}\label{5-3}
    \tan\mu=\frac{2M^2}{k^2}\sin2\eta.
\end{equation}
This expression permits one to express Eq.~(\ref{5-2}) as
\begin{equation}\label{5-4}
    r=\frac{4\theta}{k^3}\sqrt{M^2(M^2-2)\cos^2\eta+1}
\end{equation}
and Eq.~(\ref{4-5}) as
\begin{equation}\label{5-5}
    \tan\chi=\frac{(1+k^2/2)\tan\eta}{M^2-(1+k^2/2)}.
\end{equation}
At last, the curves with constant phase $\theta$ are given in Cartesian
coordinates by the formulas
\begin{equation}\label{5-6}
    \begin{split}
    &x=r\cos\chi=\frac{4\theta}{k^3}\cos\eta(1-M^2\cos2\eta),\\
    &y=r\sin\chi=\frac{4\theta}{k^3}\sin\eta(2M^2\cos^2\eta-1).
    \end{split}
\end{equation}
Thus, we have found the expressions describing the linear wave pattern
in a parametric form where the parameter $\eta$ changes in the
interval
\begin{equation}\label{5-7}
    -\arccos{\frac1M}\leq\eta\leq\arccos{\frac1M}.
\end{equation}
For $\eta=0$ we have $x<0$ and $y=0$, that is small values of
$\eta$ correspond to the wave before the obstacle. The
boundary values $\eta=\pm\arccos{(1/M)}$ correspond to the
lines
\begin{equation}\label{5-8}
    \frac{x}y=\pm\sqrt{M^2-1}=\pm a_M,
\end{equation}
that is the curves of constant phase become asymptotically the
straight lines parallel to the Mach cone. The general pattern
is shown in Fig.~2.

\section{Numerical simulations}

We have compared the above approximate analytical theory with the
exact numerical solution of the GP equation, which in
non-dimensional units (\ref{2-4}) takes the standard form
\begin{equation}\label{6-1}
    i\psi_t=-\tfrac12(\psi_{xx}+\psi_{yy})+U(x,y)\psi+|\psi|^2\psi,
\end{equation}
which corresponds for $U = 0$ to the system (\ref{2-5}) with
$$
\psi=\sqrt{n}\exp\left(i\int^{\br}\mathbf{u}(\br',t)d\br'\right).
$$
In our simulations the obstacle was modeled by an impenetrable
disk with radius $r=1$. Such an obstacle introduces large enough
perturbation into the flow to generate an oblique solitons pair
behind it (see \cite{egk1}). We assume that at the initial moment
$t=0$ there is no disturbance in the
condensate, so that it is described by the plane wave function
$$
\left.\psi(x,y)\right|_{t=0}=\exp(iMx)
$$
corresponding to a uniform condensate flow. For large enough
evolution time, the wave pattern around the obstacle tends to a
stationary structure. An example of such a structure for $M=2$ is
shown in Fig.~2.
\begin{figure}[bt]
\includegraphics[width=8cm,height=6cm,clip]{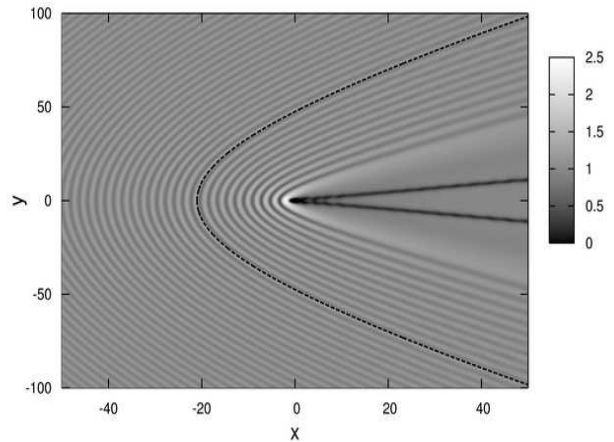}
\caption{Numerically calculated wave pattern of stationary linear
waves generated in the flow of BEC past an obstacle. The Mach
number is equal to $M=2$ and the radius of an impenetrable
obstacle to $r=1$. Dashed line
corresponds to the linear analytical theory for the line of constant
phase. It corresponds to Eqs.~(\ref{5-6}) but the curve is shifted
to 1 unit of length to the left from the center of the obstacle for better
fitting to numerics. It should be noted that this distance is
negligibly small in our theory corresponding to a point-like
obstacle. A pair of oblique dark lines behind the obstacle
correspond to spatial solitons studied in \cite{egk1}.} \label{fig3}
\end{figure}
A dashed line corresponds to the analytic theory developed in the
preceding Section; as we see it agrees very well with the numerical
results for $M=2$.

The condition (\ref{10a}) indicates that the nonlinear effects grow
up with increase of $M$. To demonstrate this explicitly, we have
compared the wavelength $\lambda$ at $y=0$ calculated using the
developed linear analytic theory with the same parameter obtained
from our full numerical simulations. According to linear theory, the
wavelength at $y=0$ (i.e. $\eta=0$) is constant and equal to
\begin{equation}\label{6-3}
    \lambda=\frac{2\pi}k=\frac{\pi}{\sqrt{M^2-1}},
\end{equation}
where we have used Eq.~(\ref{4-6}). In Fig.~4 we compare this
dependence of the wavelength $\lambda$ on the Mach number $M$ with
the results of numerical simulations at the point with $n_1\approx
0.1$. As we see, Eq.~(\ref{6-3}) is very accurate for values of $M$
satisfying the condition (\ref{10a}) and discrepancy between
analytical and numerical results slightly increases with increase
of $M$. In general, this plot confirms validity of a linear theory
in the region of its applicability.
\begin{figure}[bt]
\includegraphics[width=8cm,height=6cm,clip]{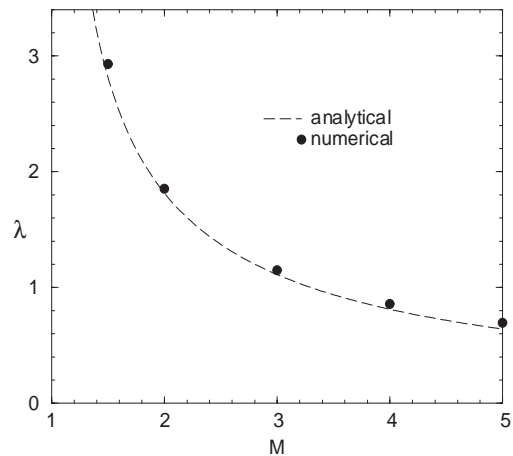}
\caption{Dependence of the wavelength of the wave generated before
the obstacle on the Mach number calculated numerically
(full time-dependent GP
equation)---dots, and analytically (linear theory)---dashed line.
} \label{fig4}
\end{figure}

\section{Conclusion}

We have developed here the theory of linear waves generated by the
flow of BEC past an obstacle. The linear approximation is correct
for small enough amplitudes of the perturbation. This condition is
satisfied in the case of small disturbance introduced by the
obstacle and not too high values of the Mach number. Our numerical
simulations confirm the analytical theory in the region of its
applicability.

\subsection*{Acknowledgements}

AMK and YGG thank RFBR (grant 05-02-17351) and AG and GAE thank
FAPESP and CNPq for financial support.

\end{document}